\documentclass[11pt,a4paper]{article}
\usepackage[margin=2.4cm]{geometry}
\usepackage[T1]{fontenc}
\usepackage{graphicx}
\usepackage{amsmath,amssymb,amsfonts}
\usepackage{xcolor}
\usepackage{booktabs}
\usepackage{float}
\usepackage{titlesec}
\usepackage[hidelinks]{hyperref}
\usepackage[numbers,sort&compress]{natbib}
\graphicspath{{./images/}}

\newcommand{\bk}{\boldsymbol{k}}
\newcommand{\bK}{\boldsymbol{K}}
\newcommand{\bu}{\boldsymbol{u}}
\newcommand{\bq}{\boldsymbol{q}}
\newcommand{\bQ}{\boldsymbol{Q}}
\newcommand{\bR}{\boldsymbol{R}}

\newcommand{\bdelta}{\boldsymbol{\delta}}
\newcommand{\bS}{\boldsymbol{S}}
\newcommand{\bD}{\boldsymbol{D}}
\newcommand{\bepsilon}{\boldsymbol{\epsilon}}

\title{\bfseries Exchange striction determines how fast antiferromagnetic insulators demagnetize}
\author{
Aleksandr Buzdakov$^{1,*}$, Ravi Kaushik$^{1}$, Nikolai Khokhlov$^{2}$,\\
Sergey Artyukhin$^{1}$, Alexey Kimel$^{2}$\\[0.6em]
\normalsize $^{1}$Istituto Italiano di Tecnologia, Via Morego 30, 16163 Genoa, Italy\\
\normalsize $^{2}$Institute for Molecules and Materials, Radboud University, 6525 AJ Nijmegen, The Netherlands\\
\normalsize $^{*}$aleksandr.buzdakov@iit.it
}
\date{}

\begin{document}
\maketitle

\begin{abstract}
Antiferromagnets combine terahertz spin dynamics with insensitivity to stray fields, and how quickly their order can be manipulated sets the speed limit on device operation.
Femtosecond optical pulses demagnetize antiferromagnetic insulators on timescales that span picoseconds to nanoseconds across compounds, and no material parameter is known that accounts for the spread or predicts where a new compound will fall. In a compensated antiferromagnet, no angular momentum needs to leave the spin system, so the rate is set by energy flow from the lattice into the spins. Time-resolved second-harmonic generation experiments show that Cr$_2$O$_3$ demagnetizes within 2~ps once the lattice is driven above the Néel temperature, two orders of magnitude faster than the structurally similar FeBO$_3$.
First-principles calculations trace the disparity to exchange striction: short Cr-Cr contacts make the exchange coupling tenfold more sensitive to atomic displacements and widen the phase space for phonon decay into magnon pairs. Spin-lattice simulations with {\it ab initio} parameters reproduce the order of magnitude of the measured ratio.
The derivative of the exchange coupling with respect to the ionic displacement thus emerges as a computable parameter that predicts how fast an insulating antiferromagnet can be demagnetized.
The results advance our understanding of ultrafast control in insulating antiferromagnets, and suggest a practical pathway to screen candidate materials for thermally assisted antiferromagnetic memory before synthesis.
\end{abstract}


\vspace{1em}
\section{Introduction}

Three decades after the seminal observation that a femtosecond laser pulse can quench roughly half of the magnetization of metallic nickel on a picosecond timescale~\cite{bigot_1996}, the field of ultrafast magnetism continues to probe the mechanisms and speed limits of spin dynamics out of equilibrium~\cite{Kirilyuk_RevModPhys2010}.
The original demagnetization was far faster than any spin-lattice relaxation rate explainable at the time, raising the question of where the angular momentum of the ordered spins goes and through which channel~\cite{chen_2025}.
The question has practical weight: laser-induced heating near the Curie point is central to heat-assisted magnetic recording~\cite{pan2009heat, seagate_mozaic4_2026}, and the ultimate speed of any thermally driven magnetic switch is set by the same couplings.

In transition-metal ferromagnets the widely accepted picture is that Elliott-Yafet spin-flip scattering between laser-heated electrons and phonons quenches the magnetization on a sub-picosecond timescale by dissipating spin angular momentum into the lattice~\cite{koopmans_2010}.
This electronic channel is unavailable in magnetic insulators, and laser-induced demagnetization of ferro- and ferrimagnetic insulators accordingly proceeds on sub-nanosecond rather than femtosecond timescales~\cite{hansteen_2006}.
Antiferromagnets occupy an intermediate position.
Because their two sublattices carry opposite magnetizations, compensated antiferromagnets can equilibrate internally through inter-sublattice exchange without transferring net angular momentum to the lattice~\cite{mentink_2012, baryakhtar_2013}, and the angular-momentum sink that constrains ferromagnetic demagnetization is therefore not required.
What sets the ultimate speed at which antiferromagnetic order can be melted in an insulator is instead the rate at which \emph{energy} flows from the laser-heated lattice into the spin subsystem.

Antiferromagnets have moved to the centre of spintronics research, through their ultrafast dynamics~\cite{windsor_2022, lee_2024, lee_2022} and through the emergence of altermagnetism.
Metallic antiferromagnets have been shown to demagnetize more rapidly than their ferromagnetic counterparts~\cite{thielemann_2017}, and insulating antiferromagnets exhibit a wide range of demagnetization times spanning picoseconds to nanoseconds~\cite{kimel_2002, kuntu_2024, khusyainov_2023, sala_2016, johnson_2015}.
The origin of this broad spread, and in particular what sets the fastest attainable timescale in a given compound, has remained unresolved.

In this work we show that the spread is governed by exchange striction, the dependence of the exchange integrals on ionic displacements.
We report the ultrafast melting of antiferromagnetic order in insulating Cr$_2$O$_3$ and compare it to FeBO$_3$, which shares the $R\bar{3}c$ space group.
At low pump fluence, the demagnetization time $\tau$ increases on approach to the N\'eel temperature, following the expected critical slowing down.
When the fluence is raised so that the final temperature exceeds $T_N$, the critical region is bypassed and $\tau$ drops below 2~ps, more than two orders of magnitude shorter than in FeBO$_3$~\cite{kimel_2002}.
Atomistic spin-dynamics simulations with a phenomenological damping reproduce the critical peak but predict a gradual decline of $\tau$ above $T_N$, whereas the measured decline is steep.
This discrepancy points to a spin-lattice coupling that grows with temperature.
First-principles calculations of the exchange-striction tensor and the one-phonon-to-two-magnon phase space account for the cross-material disparity, and coupled spin-lattice dynamics simulations based entirely on \textit{ab initio} parameters reproduce the order of magnitude of the observed ratio with no adjustable constants.

\section{Ultrafast Demagnetization of Cr$_2$O$_3$}

\begin{figure}
\centering
\includegraphics[width=0.85\linewidth]{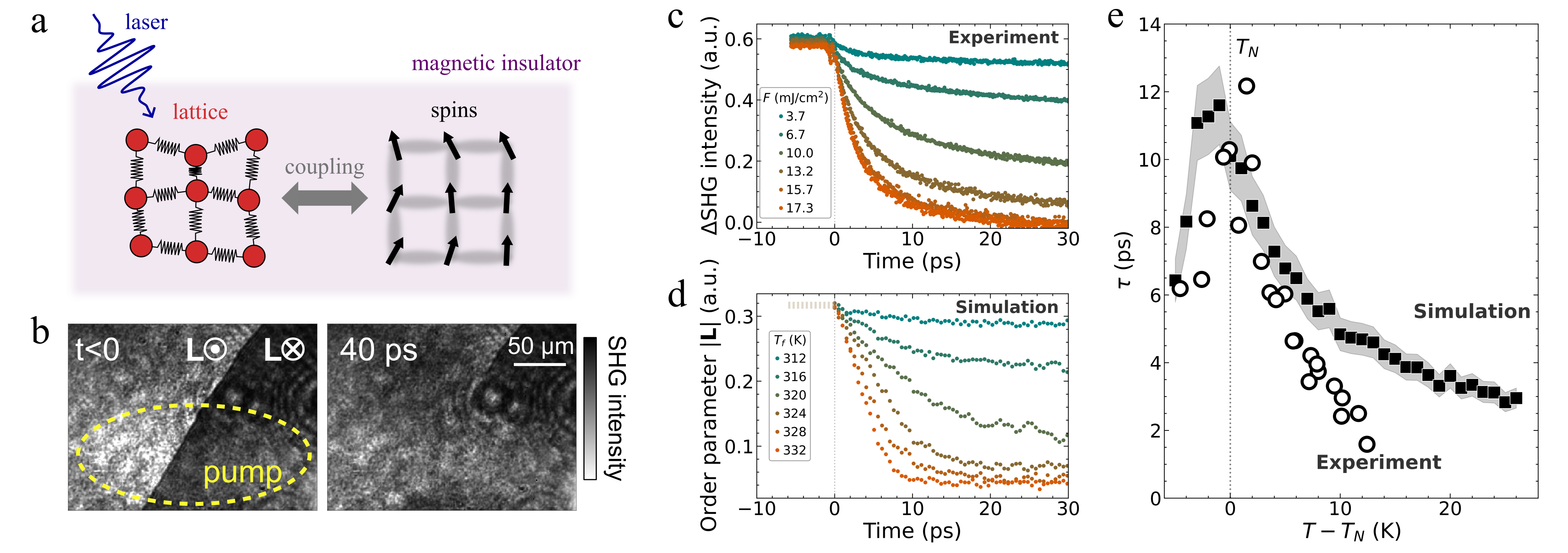}
\caption{\textbf{Ultrafast demagnetization of Cr$_2$O$_3$ and atomistic spin-dynamics simulations.}
(a) Schematic of the experiment: the pump pulse heats the lattice subsystem, and the resulting hot phonons transfer energy to the spins, reducing the order parameter $|\mathbf{L}|$.
(b) Second-harmonic-generation snapshots of a two-domain region of Cr$_2$O$_3$ before and 40~ps after a pump pulse of fluence $F=17.3$~mJ/cm$^{2}$. The domain contrast reflects the magnitude of N\'eel vector $\mathbf{L}$ and drops sharply after the pump pulse. The dashed line marks the pumped area.
(c) Time-dependent SHG contrast at pump fluences from 3.7 to 17.3~mJ/cm$^{2}$, recorded at an initial temperature of 300~K.
(d) Time evolution of the order parameter $|\mathbf{L}|$ in atomistic spin-dynamics simulations following a sudden rise of the heat-bath temperature.
(e) Characteristic demagnetization time $\tau$ as a function of $T-T_N$ for experiment (circles, $T_N=307.3$~K) and simulation (squares, $T_N^{\rm sim}=320$~K). Both datasets peak near $T_N$ due to critical slowing down. Above $T_N$ the experimental $\tau$ drops more steeply than the simulation. See Methods for the definition of $\tau$ and for simulation details.}
\label{fig:asd_and_experiment}
\end{figure}

\subsection{Sample and Experimental Setup}

We investigate ultrafast laser-induced melting of magnetic order in the compensated antiferromagnet Cr$_2$O$_3$ and compare the result with previously reported data on FeBO$_3$~\cite{kimel_2002} (Figure~\ref{fig:asd_and_experiment}a).
Neither material carries a significant net magnetization.
The $1^{\circ}$ canting in FeBO$_3$ produces a small ferromagnetic moment that is negligible for the demagnetization dynamics.
This is essential for fast melting in antiferromagnetic insulators: the sublattices equilibrate with each other and with the lattice, the order parameter shrinks as the spins heat up, and no significant angular momentum has to be transferred to the lattice.

The two compounds differ in the arrangement of their magnetic ions.
Cr$_2$O$_3$ has four Cr$^{3+}$ sites per magnetic unit cell with sublattice magnetizations $\mathbf{M}_i$ ($i=1{-}4$) ordered as $\uparrow\downarrow\uparrow\downarrow$ along [111]; FeBO$_3$ has two Fe$^{3+}$ sites ordered as $\leftarrow\rightarrow$ in the plane perpendicular to $[111]$, with boron at the remaining cation positions (Figure~\ref{fig:magnons_phonons_magnetostriction}a).
The Cr$_2$O$_3$ moments sum to zero and the order is fully described by $\mathbf{L}=\mathbf{M}_1-\mathbf{M}_2+\mathbf{M}_3-\mathbf{M}_4$.
In FeBO$_3$ the Fe$^{3+}$ spins are canted by about $1^{\circ}$ in the $(111)$ plane, so a small ferromagnetic moment coexists with $\mathbf{L}$.

Magnetic domains of opposite $\mathbf{L}$ orientation in Cr$_2$O$_3$ are imaged by second-harmonic generation (SHG) from circularly polarized probe pulses, following Fiebig et al.~\cite{fiebig1995domain}.
At 296~K, the sample exhibits a stable two-domain configuration separated by a single domain wall (Figure~\ref{fig:asd_and_experiment}b).
We probe the dynamics with a time-resolved pump-probe technique (Methods), with the pump centered on the domain wall, so that the time-dependent SHG contrast between the two domains tracks the order parameter $|\mathbf{L}|$ directly~\cite{fiebig1995domain, satoh_2007}.
Once the final temperature exceeds the N\'eel temperature $T_N=307.3$~K, the contrast vanishes together with the domain wall (Figure~\ref{fig:asd_and_experiment}b).
The pristine state is restored within 1~ms between pulses, ensuring stroboscopic reproducibility.

\subsection{Experimental Observations}

We quantify the time evolution of the magnetic order through the normalized SHG contrast $|I_\circ-I_\bullet|/(I_\circ+I_\bullet)$, where $I_\circ$ and $I_\bullet$ are the SHG intensities from the bright and dark regions of the time-delayed snapshots (Figure~\ref{fig:asd_and_experiment}b).
The transients at various pump fluences are shown in Figure~\ref{fig:asd_and_experiment}c.
We define the characteristic demagnetization time $\tau$ as the delay at which the contrast has decayed to $1/e$ of its total drop.
The same criterion is used for all transients in this work, both experimental and simulated; no functional form is fitted.
The resulting $\tau$ depends non-monotonically on the final sample temperature (Figure~\ref{fig:asd_and_experiment}e; final-temperature procedure in Methods).

At low fluences the final temperature stays below $T_N$ and $\tau$ grows on approach to $T_N$, reaching 12~ps.
This is the critical slowing down expected near a second-order transition and previously reported in antiferromagnets~\cite{Zheng_CoO_PRB2018, khusyainov_2023}: the correlation length diverges and long-wavelength modes relax slowly.
At higher fluences the final temperature crosses $T_N$ and the system bypasses the critical region.
The order parameter melts in as little as 2~ps, exposing the bare spin-lattice energy-transfer rate.
This intrinsic timescale is more than two orders of magnitude shorter than the value reported for FeBO$_3$~\cite{kimel_2002}.
Earlier SHG measurements on Cr$_2$O$_3$~\cite{sala_2016} reported sub-picosecond transients at 77~K and low fluence, but in a regime where the order parameter changes by less than 1\% and the dynamics reflect excited-state structural distortions rather than the melting of magnetic order observed here.

\subsection{Atomistic Spin-Dynamics Simulations}

To clarify the microscopic origin of this contrast, we model the response of $|\mathbf{L}|$ to a sudden phonon-bath temperature jump using atomistic spin-dynamics (ASD) simulations.
The system is equilibrated at 309~K, and at $t=0$ the bath temperature is stepped to a value between 312 and 346~K.
This protocol matches the relevant initial condition: the pump first excites electronic degrees of freedom that decay into phonons on sub-picosecond timescales~\cite{guo_2017}, leaving the lattice hotter than the spin subsystem on the timescale of the measurement.
We extract $\tau$ from the simulated transients (Figure~\ref{fig:asd_and_experiment}d) using the same $1/e$ criterion as for the experiment.

The simulated $\tau(T)$ shows a $\lambda$-shaped peak at $T_N$ and decays on both sides (Figure~\ref{fig:asd_and_experiment}e), confirming that ASD captures the critical slowing down built into the spin Hamiltonian.
Above $T_N$, however, experiment and simulation diverge.
The experimental $\tau$ falls from 10~ps to 2~ps within roughly 10~K, while the simulated $\tau$ declines gradually over more than 20~K.

The slope discrepancy points to a coupling whose effective strength grows with temperature and is missing from ASD.
In ASD the coupling to the heat bath is a constant Gilbert damping $\alpha$ by construction; changing $\alpha$ rescales the entire $\tau(T)$ curve without steepening its decline above $T_N$, and neither anisotropy nor supercell size steepens it either.
The steeper experimental trend therefore signals a spin-lattice coupling that itself grows as the lattice heats up.
We identify the interaction behind it in the next section.

\section{Magnon-Phonon Interactions in Cr$_2$O$_3$ and FeBO$_3$}

\begin{figure}
\centering
\includegraphics[width=0.85\linewidth]{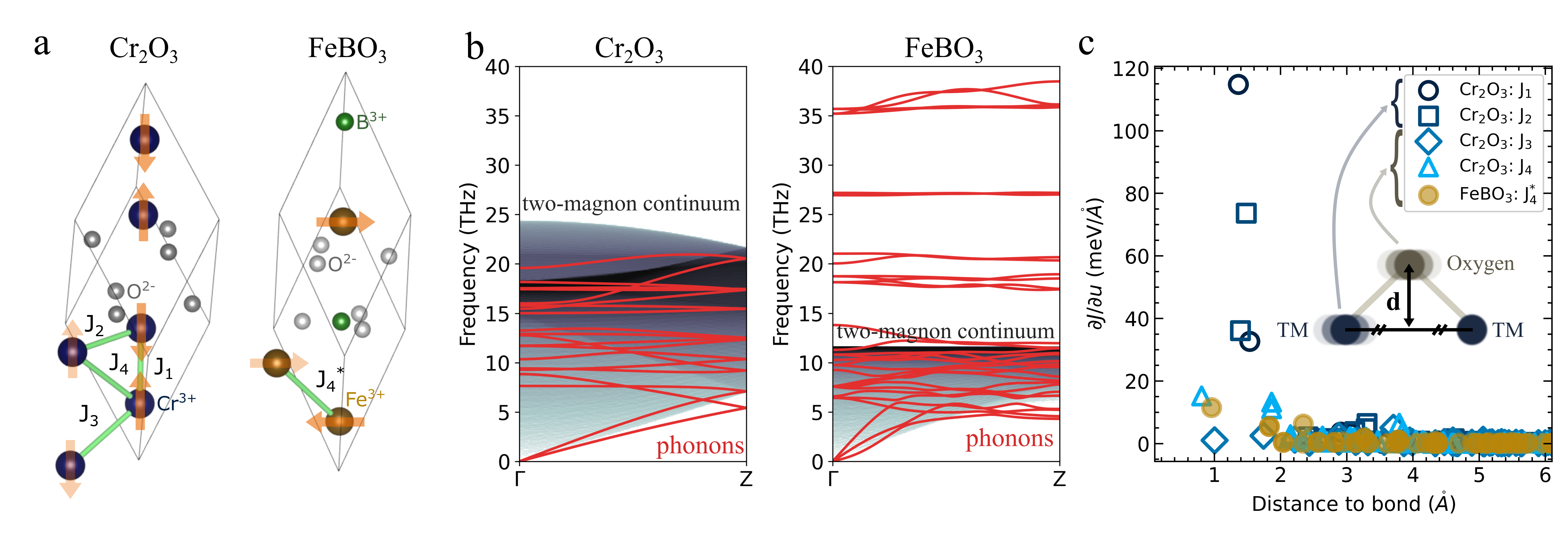}
\caption{\textbf{Magnons, phonons and exchange striction in Cr$_2$O$_3$ and FeBO$_3$.}
(a) Rhombohedral primitive cells of Cr$_2$O$_3$ (left) and FeBO$_3$ (right). Arrows indicate the sublattice magnetization, bold lines mark the symmetry-inequivalent exchange paths.
(b) Phonon dispersions (red lines) and two-magnon continuum (grey shading) along the $\Gamma$-$Z$ path. The continuum is obtained by sampling $\omega_{\mu}(\bk)+\omega_{\nu}(\bQ-\bk)$ over the full Brillouin zone for each $\bQ$ on the path. Phonon branches lying inside the continuum satisfy the resonance condition in Eq.~\eqref{eqn:golden_rule_main} and can decay into a magnon pair. The Cr$_2$O$_3$ continuum extends up to ${\sim}25$~THz and overlaps almost the entire phonon spectrum, while the FeBO$_3$ continuum is confined below 12~THz, leaving most phonon branches without a resonance partner.
(c) Exchange-striction constants $\partial J/\partial u$ for each symmetry-inequivalent bond, plotted against the atom-to-bond-centre distance $d$. In Cr$_2$O$_3$ the dominant contributions come from the Cr atoms themselves on the short $J_1$ and $J_2$ bonds and approach 100~meV/\AA. In FeBO$_3$ the dominant contributions come from the bridging oxygen on the $J_4^{*}$ bond, and no bond exceeds ${\sim}10$~meV/\AA.}
\label{fig:magnons_phonons_magnetostriction}
\end{figure}

The previous section showed that an ASD model with constant $\alpha$ misses two features of the experiment: the two-order-of-magnitude difference in $\tau$ between Cr$_2$O$_3$ and FeBO$_3$, and the steep drop of $\tau$ above $T_N$ (Figure~\ref{fig:asd_and_experiment}e).
Both call for a microscopic spin-lattice coupling that varies between materials and grows with temperature.
We show below that Heisenberg exchange striction supplies it.

We compute the Heisenberg exchange constants of both compounds from DFT$+U$ (Methods).
For Cr$_2$O$_3$ we keep interactions up to the fourth-nearest neighbour: $J_1=-12.25$~meV and $J_2=-10.07$~meV are antiferromagnetic, $J_3=+1.91$~meV and $J_4=+3.54$~meV are ferromagnetic.
The values agree with previous DFT calculations~\cite{shi_2009} and with the neutron-scattering analysis of Samuelsen \emph{et al.}~\cite{samuelsen_1970}, and reproduce the measured N\'eel temperature of 307~K within 4\% (Figure~S1, Supplementary Information).
In FeBO$_3$ the magnetic structure is dominated by a single Fe-Fe bond at 3.60~\AA~\cite{diehl_1975} with $J_4^{*}=-8.71$~meV (Figure~\ref{fig:magnons_phonons_magnetostriction}a), close to the value of ${\sim}10$~meV inferred from the canting angle and the Dzyaloshinskii-Moriya interaction~\cite{dmitrienko_2014}; the calculated N\'eel temperature is 300~K, against the measured 348~K~\cite{eibschutz_1970}.
The shorter bonds analogous to $J_1$, $J_2$ and $J_3$ in Cr$_2$O$_3$ are absent in FeBO$_3$: the calcite-type FeBO$_3$ structure places the Fe atoms on a Bravais lattice with no nearest neighbour closer than 3.60~\AA~\cite{diehl_1975}, while in the corundum-type Cr$_2$O$_3$ structure each Cr has an additional partner at 2.65~\AA{} along the body diagonal and three more at 2.89~\AA{} in the buckled basal layer.

The fourth-neighbour bonds in the two compounds carry very different exchange constants ($J_4 = +3.54$~meV in Cr$_2$O$_3$, $J_4^{*} = -8.71$~meV in FeBO$_3$) despite a similar M-O-M geometry, because the underlying mechanisms differ.
In FeBO$_3$ the Fe$^{3+}$ ($d^5$) configuration leaves every $d$ orbital half-filled, and oxygen superexchange through the open $e_g$-O-$e_g$ channel produces a strong antiferromagnetic bond following the Goodenough-Kanamori rules~\cite{goodenough_1963, kanamori_1959}, consistent with the value inferred from the canting and the Dzyaloshinskii-Moriya interaction~\cite{dmitrienko_2014}.
In Cr$_2$O$_3$, by contrast, oxygen-mediated superexchange is largely ineffective.
Shi \emph{et al.}~\cite{shi_2009} demonstrated this by artificially shifting the oxygen $p$ states in DFT and observing that the magnetic energies are unchanged: the magnetic energetics of Cr$_2$O$_3$ are dominated by direct $t_{2g}$-$t_{2g}$ overlap between Cr ions.
The strong $e_g$-O-$e_g$ superexchange channel of FeBO$_3$ is closed because the $e_g$ subshell is empty in Cr$^{3+}$ ($d^3$), and the residual $t_{2g}$-O-$t_{2g}$ channel has weak orbital overlap.
The strong $J_1$ and $J_2$ at 2.65 and 2.89~\AA{} therefore arise from direct overlap; the longer-range $J_3$ and $J_4$ are weak by comparison ($|J_4|<4$~meV).

Magnons are computed within linear spin-wave theory; phonons separately within DFT$+U$ without spin-orbit coupling (red lines for phonons in Figure~\ref{fig:magnons_phonons_magnetostriction}b; full dispersions for magnons and phonons in Figure~S2, Supplementary Information).

The coupling between magnons and phonons has two contributions: Heisenberg exchange striction, from the displacement dependence of $J_{ij}$, and Dzyaloshinskii-Moriya striction, from the displacement dependence of $\bD_{ij}$.
Both are derived in Supplementary Information.
Because $\hat{\bS}_i\cdot\hat{\bS}_j$ is quadratic in magnon operators at leading order, exchange striction drives a \emph{one-phonon--two-magnon} process: a phonon at momentum $\bq$ decays into a magnon pair with momenta $\bk$ and $\bq-\bk$.
The DM channel is allowed by symmetry but its matrix elements scale as $|\partial\bD/\partial\bu|^{2}$ rather than $|\partial J/\partial\bu|^{2}$, giving a $(D/J)^{2}$ suppression that makes it negligible (Supplementary Information).
We focus on the exchange channel, with rate 
\begin{equation}\label{eqn:golden_rule_main}
  \tau_{\bq,\lambda}^{-1}
  = \frac{2\pi}{\hbar N} \sum_{\bk,\mu,\nu}
    |V_{\bk,\bq}^{\lambda}|^{2} \,
    \delta\bigl(\hbar\Omega_{\bq,\lambda}
          - \epsilon_{\bk,\mu} - \epsilon_{\bq-\bk,\nu}\bigr),
\end{equation}
where $V_{\bk,\bq}^{\lambda}$ is the exchange-striction vertex defined in Supplementary Information, $\Omega_{\bq,\lambda}$ is the phonon frequency at momentum $\bq$ in branch $\lambda$, $\epsilon_{\bk,\mu}$ is the magnon energy at $\bk$ in branch $\mu$, and $N$ is the number of unit cells.

The phonon-magnon kinematics differ markedly between the two materials.
Figure~\ref{fig:magnons_phonons_magnetostriction}b shows the dispersions and the two-magnon continuum along the rhombohedral high-symmetry path $\Gamma$-$Z$, which runs along the body diagonal of the unit cell.
In Cr$_2$O$_3$ the strong $J_1$ and $J_2$ exchanges produce a wide magnon band, and the resulting two-magnon continuum extends up to 25~THz, overlapping almost the entire phonon spectrum.
In FeBO$_3$ the single antiferromagnetic exchange yields a narrower magnon band, and the continuum is confined below 12~THz; only the lower phonon branches fall inside it, while most phonons sit at higher energies and have no resonance partner.

Since laser excitation populates phonons throughout the Brillouin zone, the rate in Eq.~\eqref{eqn:golden_rule_main} must be integrated over all $\bq$.
We compute the kinematic phase space of the one-phonon-two-magnon channel from the \emph{ab initio} dispersions on a regular grid spanning the full BZ (Methods; Figures~S3 and~S4, Supplementary Information).
The local picture along $\Gamma$-$Z$ holds in the BZ average: the Cr$_2$O$_3$ continuum overlaps almost all phonon branches, providing decay channels across the full spectrum, while in FeBO$_3$ the overlap is restricted to the low-frequency phonons.
The integrated phase space is about three times larger in Cr$_2$O$_3$.
The one-phonon-one-magnon phase space, restricted to phonon-magnon band crossings, is of similar magnitude in the two materials and several times smaller than the two-magnon channel in both (Supplementary Information), so the cross-material asymmetry is concentrated in the two-magnon process.

The exchange-striction matrix elements amplify this asymmetry, but they do so for a reason different from the kinematics.
Figure~\ref{fig:magnons_phonons_magnetostriction}c shows $|\partial J/\partial u|$ as a function of the atom-to-bond-centre distance for both compounds (Methods).
In Cr$_2$O$_3$ the $J_1$ and $J_2$ bonds give $|\partial J/\partial u|\sim 100$~meV/\AA, with the dominant contribution coming from displacements of the Cr atoms themselves rather than of the bridging oxygen.
This reflects a second peculiarity of $J_1$ and $J_2$ in Cr$_2$O$_3$: at the short Cr-Cr distances of 2.65 and 2.89~\AA{} the magnetic interaction is dominated by direct $t_{2g}$-$t_{2g}$ overlap between Cr ions rather than by oxygen-mediated superexchange~\cite{shi_2009}, and direct exchange depends exponentially on the cation-cation distance.
Small Cr displacements then produce large changes in $J$.
In FeBO$_3$ no such short cation-cation contacts exist; the only available channel is oxygen-mediated superexchange, modulated by displacements of the bridging oxygen, which yields $|\partial J/\partial u|\lesssim 10$~meV/\AA.
The factor $|V|^{2}$ in Eq.~\eqref{eqn:golden_rule_main} therefore favours Cr$_2$O$_3$ by roughly two orders of magnitude.

These two factors enter Eq.~\eqref{eqn:golden_rule_main} multiplicatively.
Phase space gives a factor of three, the matrix element a factor of about $10^{2}$, and the predicted rate ratio is ${\sim}3\times 10^{2}$.
This is an upper bound: branch-dependent striction, polarization projections, and finite-temperature renormalization of the magnon spectrum all reduce it.
The measured ratio of order $10^{2}$ between the two materials is consistent with this estimate and identifies exchange striction as the dominant channel.

The same coupling is consistent with the steep $\tau(T)$ slope above $T_N$ in Figure~\ref{fig:asd_and_experiment}e.
At finite temperature the rate in Eq.~\eqref{eqn:golden_rule_main} acquires Bose factors $[1+n(\omega)]$, which increase the energy flow from the heated lattice.
More importantly, with $|\partial J/\partial u|\sim 100$~meV/\AA{} and room-temperature thermal displacements of order 0.05~\AA, the instantaneous exchange is modulated by several meV, a sizeable fraction of $J_1$ itself: the spin-lattice coupling in Cr$_2$O$_3$ is not a small perturbation, and a constant phenomenological damping cannot capture how it grows with temperature.
The spin-lattice dynamics simulations of the next section include this modulation nonperturbatively.

\section{Coupled Spin-Lattice Dynamics}

The previous section estimated the rate ratio between Cr$_2$O$_3$ and FeBO$_3$ from Fermi's golden rule.
That estimate is perturbative in the spin-lattice coupling and uses zero-temperature dispersions.
To test whether the full nonlinear, finite-temperature dynamics confirms the picture, we carry out coupled spin-lattice dynamics (SLD) simulations for both materials~\cite{hellsvik_2019, miranda_2025}.

The method integrates the Landau-Lifshitz-Gilbert equation for the spins together with Newton's equations for the ions.
The spin subsystem is governed by the exchange tensor $J^{\alpha\beta}_{ij}$, the lattice subsystem by the force-constant tensor $\Phi^{\mu\nu}_{kl}$, and the coupling between them by the exchange-striction tensor $\partial J^{\alpha\beta}_{ij}/\partial u^{\mu}_{k}$.
All three tensors come from the DFT$+U$ calculations of the previous section; no parameter is adjusted.

\begin{figure}[htbp]
\centering
\includegraphics[width=0.6\linewidth]{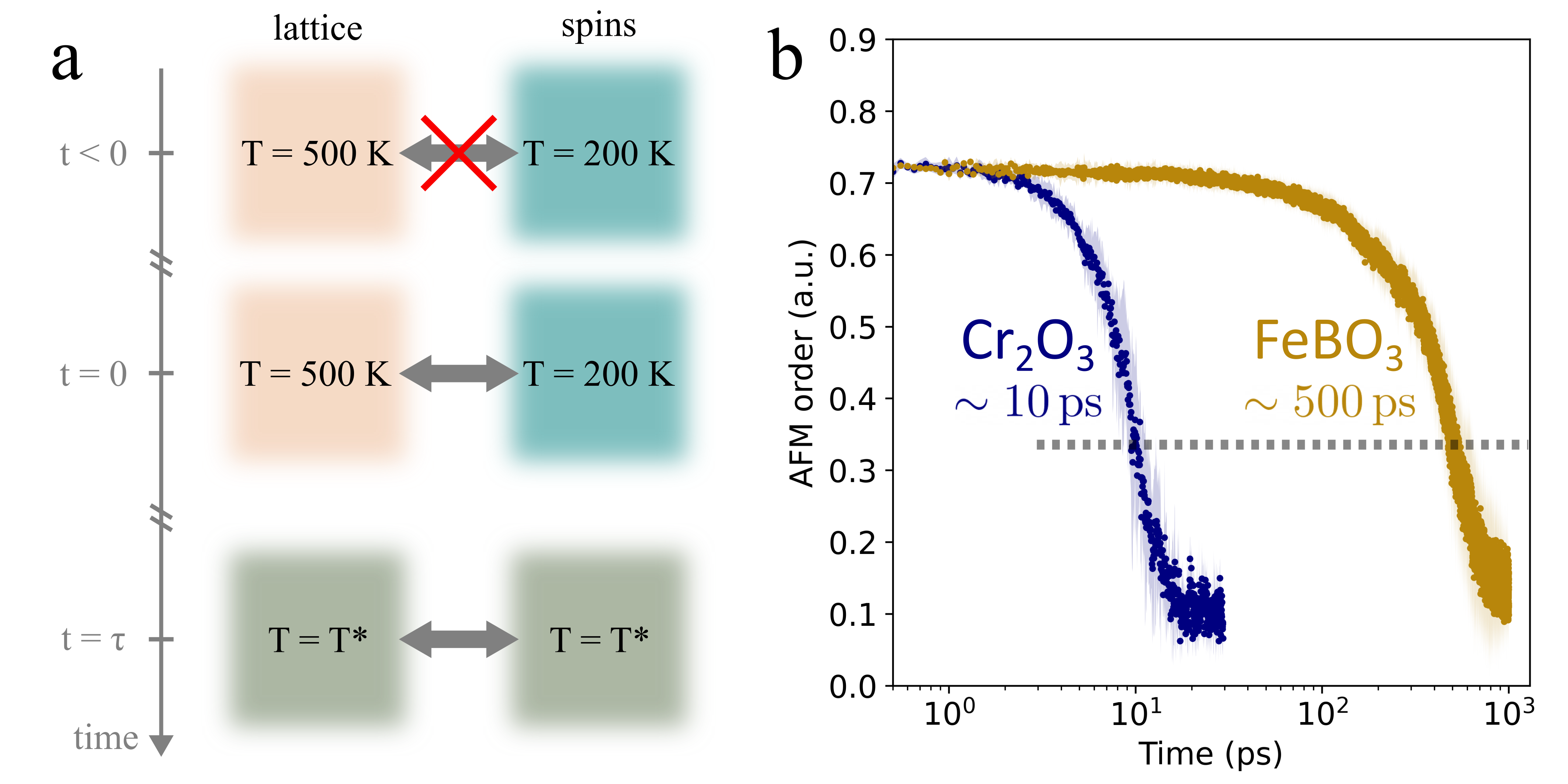}
\caption{\textbf{Coupled spin-lattice dynamics simulations.}
(a) Simulation protocol. For $t<0$ the lattice is thermalized at 500~K and the spin subsystem at 200~K with the spin-lattice coupling switched off (red cross), so the two subsystems reach equilibrium independently. At $t=0$ the coupling is turned on (grey arrow) and the damping is removed; the combined system evolves microcanonically while energy flows from the hot lattice into the spins until the two subsystems reach a common effective temperature $T^{*}$ at $t=\tau$.
(b) Time evolution of the order parameter $|\mathbf{L}|$ for Cr$_2$O$_3$ (blue) and FeBO$_3$ (gold), shown on a logarithmic time axis. The starting value below unity reflects the finite spin temperature of 200~K. Symbols are averages over independent trajectories; shaded bands give the $1\sigma$ spread. The dashed grey line marks the $1/e$ level of the total drop, defining the relaxation time $\tau$. We obtain $\tau\approx 10$~ps for Cr$_2$O$_3$ and $\tau\approx 500$~ps for FeBO$_3$.}
\label{fig:SLD}
\end{figure}

The protocol is shown in Figure~\ref{fig:SLD}a.
The lattice is thermalized at 500~K and the spin system at 200~K, with the spin-lattice coupling switched off so the two subsystems reach equilibrium independently.
This temperature imbalance mimics the post-excitation state in which the pump pulse has heated the lattice well above the cold spin bath.
At $t=0$ we switch the coupling on and remove the damping; the combined system then evolves microcanonically while energy flows from the hot lattice into the spins until the two subsystems reach a common effective temperature.
We monitor this equilibration through the antiferromagnetic order parameter (Figure~\ref{fig:SLD}b).
The order parameter starts below unity because the spin subsystem is already at 200~K, and does not decay all the way to zero because the supercell cannot host the long-wavelength fluctuations that fully disorder the N\'eel state; the relevant observable is the relaxation time, not the asymptote.

The two materials relax on very different timescales: $\tau_{\mathrm{Cr}_{2}\mathrm{O}_{3}}\approx 10$~ps and $\tau_{\mathrm{FeBO}_{3}}\approx 500$~ps, a factor of 50.
This sits within the golden-rule upper bound of $\sim 3\times 10^{2}$ from the previous section, and the measured ratio of order $10^{2}$ falls between the SLD result and that bound.
DM contributions enter at $(D/J)^{2}$ in the matrix elements (Supplementary Information) and do not affect the comparison at leading order.

Most spin-dynamics simulations treat the exchange integrals as fixed parameters and ignore their dependence on ionic positions.
Recent work has relaxed this approximation~\cite{strungaru_2021, hellsvik_2019, miranda_2025}; our simulations extend this line, so far applied mainly to elemental metals, to multi-sublattice antiferromagnetic insulators, and show that once the displacement dependence of $J_{ij}$ is included, a parameter-free first-principles model reproduces the experimental order of magnitude.

\section{Conclusions}
In compensated antiferromagnets, no angular momentum needs to leave the spin system, and magnetic order therefore melts only as fast as energy flows from the heated lattice to the spins.
Exchange striction carries this flow.
Short Cr-Cr contacts make the exchange coupling of Cr$_2$O$_3$ tenfold more sensitive to atomic displacements than that of FeBO$_3$ and widen the phase space for phonon decay into magnon pairs. Together, the two factors account for the two-orders-of-magnitude difference in demagnetization time.
Spin-lattice simulations built entirely on \emph{ab initio} parameters reproduce the order of magnitude of the measured ratio.

The rate thus rests on three ingredients: the striction $\partial J/\partial u$, the thermal displacements that multiply it, and the overlap of the two-magnon continuum with the phonon spectrum.
All three follow from first-principles calculations, so the demagnetization speed of an insulating antiferromagnet can be predicted before the experiment.
Strain, pressure, or substitution that shortens exchange-carrying cation-cation bonds raises it; magnets coupled only through oxygen-mediated superexchange, such as calcite-structure MnCO$_3$, are at the slow end.

The same tensor quantifies the coupling for coherent control.
It underlies the proposed phononic switching protocol in Cr$_2$O$_3$~\cite{fechner_2018} and the demonstrated phononic modification of exchange interactions~\cite{afanasiev_2021}. The magnon-phonon Fermi resonance realized in CoF$_2$~\cite{mashkovich_2021,metzger_2024} is the coherent counterpart of the incoherent one-phonon-two-magnon channel identified here.

The same three ingredients rank the systems to try next.
Van der Waals antiferromagnets score on the displacements: their soft lattices amplify the exchange modulation even where the striction itself is moderate.
Insulating altermagnets such as MnTe inherit the energy-flow bottleneck unchanged, since they too lack conduction electrons, and which interaction carries the flow there is an open question.
In metallic antiferromagnets, hot carriers add a parallel channel, and its competition with the lattice one lies beyond the present framework.
For the insulators, the spread of demagnetization times is no longer an empirical fact: it is a computable material property.

\section{Methods}

\subsection{Sample and Optical Setup}

For the experimental study, we used a single crystal of Cr$_2$O$_3$ commercially available (MaTecK) with dimensions of 5$\times$5$\times$0.5~mm$^3$, a surface orientation (111), polished from both sides with roughness $<0.01$~$\mu$m and orientation precision $<0.1^{\circ}$.
The sample was placed on a heater to control its temperature.
Measurements were performed in the absence of external magnetic and electric fields.

The experimental geometry is shown in Figure~S5a, Supplementary Information.
The 40~fs pulses with a central wavelength of 800~nm at a repetition rate of 1~kHz, generated using a Ti:sapphire amplifier, were divided into pump and probe paths.
The probe pulse was converted to 1200~nm in an optical parametric amplifier and circularly polarized with a quarter-wave plate before illuminating a 4~mm$^2$ spot on the sample.
The probe fluence was 2~mJ/cm$^2$.
The SHG probe signal from the sample at 600~nm passed through a bandpass filter before reaching the charge-coupled device (CCD) camera with a micro-objective lens.
The pump pulses were frequency-doubled to 400~nm with a beta barium borate crystal.
The resulting $p$-polarized pump pulses were focused on the sample from the camera side with an oblique incidence of 40$^\circ$ in an elliptical spot with a long axis of 200~$\mu$m.
To achieve temporal resolution, the probe pulse passed through a mechanical delay line with a retroreflector.
The dielectric permittivity of the sample at 400~nm (Figure~S5b, Supplementary Information) was measured by variable-angle spectroscopic ellipsometry (VASE, J.A.~Woollam Co.).

\subsection{Final Temperature Calculation}

The pump pulse deposits energy into the lattice according to a Beer-Lambert absorption profile.
We compute the resulting depth-dependent temperature rise as follows.
The thermal energy density stored in the lattice at temperature $T$ is
\begin{equation}
  E_{\rm th}(T)
  = \frac{\rho}{M} \int_{T_{\rm ref}}^{T} c_p(T')\,\mathrm{d}T',
\end{equation}
where $\rho = 5220$~kg\,m$^{-3}$ is the mass density, $M = 0.152$~kg\,mol$^{-1}$ is the molar mass, and $c_p(T)$ is the molar specific heat capacity taken from Ref.~\cite{gurevich_2009}, which exhibits a $\lambda$-shaped anomaly at $T_N$.

The local energy density deposited by the pump at depth $z$ is
\begin{equation}\label{eqn:Eabs}
  E_{\rm abs}(z)
  = \frac{F\,A}{\delta_{\rm pump}}\,e^{-z/\delta_{\rm pump}},
\end{equation}
where $F$ is the incident pump fluence averaged over the pumped area, $\delta_{\rm pump}$ is the optical penetration depth at 400~nm, and $A = 1 - R$ is the absorptance, with $R$ the reflectance at the pump wavelength.
The dielectric permittivity $\varepsilon = 6.4$ of the sample at 400~nm gives an absorptance $A = 0.89$, calculated via the Fresnel formula for $p$-polarized light at $40^\circ$ incidence.
The penetration depth $\delta_{\rm pump} = 1.12$~$\mu$m is obtained from the absorption coefficient of bulk single-crystal Cr$_2$O$_3$ at 400~nm reported in Ref.~\cite{mcclure_1963}.

After the pulse, the total energy density at depth $z$ is $E_{\rm tot}(z) = E_{\rm th}(T_i) + E_{\rm abs}(z)$, with $T_i$ the initial sample temperature.
The local final temperature $T_f(z)$ is obtained by inverting the monotonic function $E_{\rm th}$: $T_f(z) = E_{\rm th}^{-1}\bigl[E_{\rm tot}(z)\bigr]$.

The measured signal is the second harmonic at 600~nm generated from the 1200~nm probe.
The 1200~nm probe lies far below the optical band gap of Cr$_2$O$_3$ (3.2~eV), so its intensity is essentially uniform across the heated layer.
The domain contrast is the interference term between the crystallographic and magnetic SHG contributions and is therefore linear in the magnetic second-harmonic field.
The field generated at depth $z$ is attenuated on the way to the surface as $\exp(-\alpha_{600}\,z/2)$, with $\alpha_{600}$ the intensity absorption coefficient at 600~nm from Ref.~\cite{mcclure_1963}, so the contrast weights depths as $\exp(-z/\delta_{\rm meas})$ with $\delta_{\rm meas} = 2/\alpha_{600} = 1.42$~$\mu$m.
The depth-resolved final temperature observed via SHG is
\begin{equation}\label{eqn:Tavg}
  \langle T_f \rangle
  = \frac{\displaystyle\int_0^{\infty} T_f(z)\,e^{-z/\delta_{\rm meas}}\,\mathrm{d}z}
         {\displaystyle\int_0^{\infty} e^{-z/\delta_{\rm meas}}\,\mathrm{d}z},
\end{equation}
which gives greater weight to the near-surface region where the probe sensitivity is highest.
In practice the integrals are evaluated numerically on a grid extending to $5\,\delta_{\rm meas}$, beyond which the exponential weight is negligible.
The transients entering Figure~\ref{fig:asd_and_experiment}e were recorded at initial temperatures $T_i = 300$, 302, 304 and 306~K for each of the six pump fluences, and $\langle T_f \rangle$ was computed separately for every $(T_i, F)$ combination.

\subsection{First-Principles Calculations}

All density-functional calculations were performed with Quantum ESPRESSO~\cite{Giannozzi_2009, Giannozzi_2017, Giannozzi_2020, QuantumEspresso}.
Electronic correlations on the transition-metal $3d$ shells were treated within the simplified rotationally-invariant DFT$+U$ scheme of Dudarev~\cite{dudarev_1998}, with a single effective parameter $U_{\rm eff}=U-J$.
Pseudopotentials were taken from version 1.3.0 of the Standard Solid State Pseudopotentials (SSSP) library, Precision protocol~\cite{prandini_2018}: Fe and O from pslibrary 0.3.1~\cite{kucukbenli_2014}, Cr from pslibrary 1.0.0~\cite{dalcorso_2014}, and B from GBRV 1.2~\cite{garrity_2014}.
Plane-wave and charge-density cutoffs were set to the SSSP-recommended values for each element, taking the maximum across species in each unit cell.

The Hubbard parameter for each compound was computed from first principles using density-functional perturbation theory in the linear-response formulation of Cococcioni and de Gironcoli~\cite{cococcioni_2005}, as implemented in the HP module of Quantum ESPRESSO~\cite{timrov_2022}.
The calculation was performed on the relaxed equilibrium structure with a $2\times 2\times 2$ q-point grid for the monochromatic perturbations, applied to each magnetic site in the primitive cell.
This yielded $U_{\rm eff}=6.06$~eV for Cr in Cr$_2$O$_3$ and $U_{\rm eff}=5.04$~eV for Fe in FeBO$_3$.
With these values the calculated band gaps are 3.39 and 2.36~eV for Cr$_2$O$_3$ and FeBO$_3$, respectively.

The exchange constants $J_{ij}$ and their displacement derivatives $\partial J_{ij}/\partial \bu_k$ were obtained from the four-state method~\cite{fourspin11}.
For each symmetry-inequivalent bond $(i,j)$ we compute total energies $E_{1\text{-}4}$ and Hellmann-Feynman forces $\boldsymbol{F}^k_{1\text{-}4}$ on every atom $k$ in four collinear spin configurations differing only in the orientations of $S_i$ and $S_j$ ($\uparrow\downarrow$, $\uparrow\uparrow$, $\downarrow\downarrow$, $\downarrow\uparrow$).
The exchange constant and its derivative follow from
\begin{align}\label{eqn:fourstate_J}
J_{ij} &= \frac{E_1-E_2-E_3+E_4}{4}, \\
-\frac{\partial J_{ij}}{\partial \bu_k} &= \frac{\boldsymbol{F}^k_1-\boldsymbol{F}^k_2-\boldsymbol{F}^k_3+\boldsymbol{F}^k_4}{4}.
\end{align}
The constants obtained this way are energies per bond of unit spin vectors, the normalization used by the atomistic and spin-lattice simulations throughout this work; in this convention negative values denote antiferromagnetic bonds.
They map onto the spin-operator normalization of Supplementary Information as $\tilde{J}=-S^{2}J$, with $S=3/2$ for Cr$^{3+}$ and $S=5/2$ for Fe$^{3+}$.
The four-state combination cancels contributions to the total energy and forces that are not associated with the $i$-$j$ bond, isolating the magnetic interaction of interest.

Phonon dispersions were computed within DFT$+U$ using the finite-displacement method as implemented in phonopy~\cite{phonopy-phono3py-JPCM, phonopy-phono3py-JPSJ}, with the magnetic ground-state ordering imposed throughout and no spin-orbit coupling.
We used a $2\times 2\times 2$ supercell of the rhombohedral primitive cell for Cr$_2$O$_3$ and a $3\times 3\times 2$ supercell for FeBO$_3$.

\subsection{Atomistic Spin-Dynamics Protocol}

The temperature dependence of the antiferromagnetic order parameter was determined both numerically and experimentally.
Numerical estimates were obtained from atomistic spin-dynamics simulations using the UppASD package~\cite{skubic2008method_UppASD, eriksson2017atomistic_UppASD}, with the exchange constants of the previous subsection as input.
The order parameter was sampled in thermal equilibrium at each temperature and the specific heat $C_V$ was computed from energy fluctuations in the same simulations.
Figure~S1a,b, Supplementary Information, show the resulting curves for Cr$_2$O$_3$ and FeBO$_3$.
The peak of $C_V$ locates the calculated N\'eel temperature at 320~K for Cr$_2$O$_3$ and 300~K for FeBO$_3$, against measured values of 307~K and 348~K.
The same UppASD framework was used for the temperature-jump runs in Figure~\ref{fig:asd_and_experiment}d,e: the system was equilibrated at 309~K and at $t=0$ the bath temperature was stepped to a final value between 312 and 346~K, with a constant Gilbert damping parameter $\alpha=5\times10^{-3}$ chosen to match typical literature values for transition-metal oxides.
Trajectories were averaged over independent initial conditions to suppress thermal noise.
The characteristic time $\tau$ is the delay at which the order parameter has decayed to $1/e$ of its total drop; no functional form is fitted.

The experimental order parameter was extracted from the SHG contrast between the two antiferromagnetic domains and is shown in Figure~S1c, Supplementary Information.
A fit to $\Delta_{\rm SHG}(T) = A\,(T_N-T)^{\beta} + y_0$ yields $T_N=307.3$~K and a critical exponent $\beta=0.54\pm 0.08$.
The non-zero offset $y_0$ above $T_N$ reflects the inhomogeneous spatial profile of the pump pulse, not residual order.

\subsection{Phonon-Magnon Kinematics}

Magnon dispersions were computed using linear spin-wave theory as implemented in the Sunny.jl package~\cite{dahlbom2025sunnyil}, with the exchange constants $J_1, J_2, J_3, J_4$ for Cr$_2$O$_3$ and $J_4^{*}$ for FeBO$_3$ obtained as described above.
The rhombohedral primitive cell of Cr$_2$O$_3$ contains four Cr atoms and gives four magnon branches (Figure~S2a, Supplementary Information); FeBO$_3$ has two Fe atoms per primitive cell and the dispersion accordingly has two branches (Figure~S2b, Supplementary Information).
The phonon spectrum of Cr$_2$O$_3$ extends to ${\sim}25$~THz, whereas that of FeBO$_3$ reaches ${\sim}38$~THz; the higher upper bound in FeBO$_3$ reflects the presence of light boron atoms, whose vibrational modes form a separate band of optical phonons well above the Cr$_2$O$_3$ spectrum.

The magnon-phonon coupling for a collinear two-sublattice antiferromagnet is derived in Supplementary Information.
After Holstein-Primakoff bosonization of the spins and standard expansion of the displacements over phonon modes, the exchange-striction interaction reduces to a one-phonon-two-magnon vertex
\begin{equation}\label{eqn:V_methods}
  V^{\lambda}_{\bk,\bq}
  = S\sum_{\bdelta}\sum_{n,\bR'}
    \sqrt{\frac{\hbar}{2 m_{n} \Omega_{\bq,\lambda}}}\;
    \bigl(\bK^{n}_{\bdelta}(\bR')\cdot\bepsilon^{n}_{\bq,\lambda}\bigr)\,
    e^{i\bq\cdot\bR'}\,e^{-i(\bq-\bk)\cdot\bdelta},
\end{equation}
where $\bK^{n}_{\bdelta}(\bR')=\partial J_{\bdelta}/\partial \bu_{n,\bR'}$ is the striction tensor for the bond at the origin with $\bdelta$ the inequivalent neighbour vector, $\bepsilon^{n}_{\bq,\lambda}$ is the phonon polarization at atom $n$, $m_n$ its mass, and $S$ the spin magnitude.
The Dzyaloshinskii-Moriya channel contributes both one-phonon-one-magnon and one-phonon-two-magnon vertices, but both are suppressed at the matrix-element level by $(D/J)^{2}$ relative to exchange striction (Supplementary Information), so we retain only the exchange-driven 1p2m contribution.

The kinematic factor in Eq.~\eqref{eqn:golden_rule_main} is evaluated from the \emph{ab initio} dispersions on a uniform $N_g\times N_g\times N_g$ grid spanning the rhombohedral Brillouin zone ($N_g=50$, reduced coordinates in $[0,1)^{3}$), so that $N=N_g^{3}$ in Eq.~\eqref{eqn:golden_rule_main}.
The energy $\delta$-function is smeared by a normalized Gaussian of width $\sigma=0.25$~THz, small compared with the phonon and magnon bandwidths but large enough to bridge the typical energy spacing between adjacent grid points.
The one-phonon-one-magnon and one-phonon-two-magnon phase-space densities (obtained by setting the matrix elements to unity in the two golden-rule rates) are shown for both materials in Figures~S3 and~S4, Supplementary Information.
Normalized per internal momentum, as in Eq.~\eqref{eqn:golden_rule_main}, the 1p2m density exceeds the 1p1m density by a factor of 3--4 in both materials, and is about three times larger in Cr$_2$O$_3$ than in FeBO$_3$; the dominant suppression of the Dzyaloshinskii-Moriya channels is the matrix-element factor $(D/J)^{2}$ (Supplementary Information).

\section*{Author Contributions}

All authors contributed to the discussion of the results.
\textbf{A.B.}: Conceptualization, Investigation (atomistic spin dynamics, spin-lattice dynamics), Formal analysis, Writing -- original draft, Visualization.
\textbf{R.K.}: Investigation (first principles calculations), Formal analysis, Writing -- review and editing.
\textbf{N.Kh.}: Investigation (experiment), Methodology (experiment), Writing -- review and editing.
\textbf{S.A.}: Conceptualization, Supervision, Methodology, Writing -- review and editing, Project Administration.
\textbf{A.K.}: Conceptualization, Supervision, Writing -- review and editing, Project Administration.

\section*{Acknowledgments}
The authors thank V. Radovskaia for ellipsometry of the sample and Dr. K. Saeedi Ilkhchy and C. Berkhout for their technical assistance with the experiment.
N.Kh. and A.K. disclose support for the research of this work from the European Research Council, ERC Grant Agreement No. 101054664 (SPARTACUS), and from the research program ``Materials for the Quantum Age'' (QuMat).

\section*{Conflicts of Interest}

The authors declare no conflicts of interest.

\section*{Data Availability Statement}

The data that support the findings of this study are available from the corresponding author upon reasonable request. All density-functional, spin-dynamics, and spin-lattice-dynamics calculations use open-source packages cited throughout the paper; custom analysis scripts are available from the corresponding author upon reasonable request.

\clearpage
\appendix
\setcounter{figure}{0}
\setcounter{equation}{0}
\renewcommand{\thefigure}{S\arabic{figure}}
\renewcommand{\theequation}{S\arabic{equation}}
\renewcommand{\thesection}{S\arabic{section}}

\begin{center}
{\Large\bfseries Supplementary Information}
\end{center}
\section{Magnon-phonon coupling from exchange striction}

This note derives the magnon-phonon vertices generated by the displacement dependence of the magnetic interactions in a collinear two-sublattice antiferromagnet. The exchange channel produces a one-phonon-two-magnon process. The Dzyaloshinskii-Moriya channels are estimated and shown to be negligible. The quantitative analysis of the main text uses the vertex~\eqref{eqn:V_ex_SI}, the rate~\eqref{eqn:golden_rule_SI}, and the kinematic densities~\eqref{eqn:D_1p1m_SI} and~\eqref{eqn:D_1p2m_SI}.

\subsection{Model and conventions}

We consider spins of magnitude $S$ on two sublattices, with sublattice $A$ ordered along $+\hat{z}$ and sublattice $B$ along $-\hat{z}$. The bare spin Hamiltonian contains Heisenberg exchange and a Dzyaloshinskii-Moriya (DM) term~\cite{moriya_1960}:
\begin{equation}\label{eqn:Hspin_SI}
  \hat{H}_{\rm spin}
  = \frac{1}{2}\sum_{i\neq j}
    \Bigl[
      J_{ij}\,\hat{\bS}_{i}\cdot\hat{\bS}_{j}
      + \bD_{ij}\cdot(\hat{\bS}_{i}\times\hat{\bS}_{j})
    \Bigr],
\end{equation}
with $J_{ij}>0$ on the dominant bonds and $|\bD_{ij}|\ll J_{ij}$.

Two conventions differ between this note and the numerical work, and we state the map once. Within this note a positive $J$ denotes an antiferromagnetic bond. The main text and the atomistic simulations use the opposite sign convention and quote energies per bond of unit spin vectors, so the published constants are $\tilde{J}_{ij} = -S^{2} J_{ij}$, and the striction tensors carry the same normalization, $\partial\tilde{J}_{ij}/\partial\bu = -S^{2}\,\partial J_{ij}/\partial\bu$. All expressions below are written for spin operators and are converted to the published values through this map.

\subsection{Holstein-Primakoff bosonization}

In each unit cell at lattice position $\bR$ we introduce Holstein-Primakoff (HP) bosons $\hat{a}_{\bR}$ for the $A$ sublattice and $\hat{b}_{\bR}$ for the $B$ sublattice:
\begin{align}\label{eqn:HP_SI}
  \hat{S}^{z}_{A,\bR} &= S - \hat{a}^{\dag}_{\bR}\hat{a}_{\bR},
  & \hat{S}^{+}_{A,\bR} &\simeq \sqrt{2S}\,\hat{a}_{\bR},
  & \hat{S}^{-}_{A,\bR} &\simeq \sqrt{2S}\,\hat{a}^{\dag}_{\bR}, \nonumber\\
  \hat{S}^{z}_{B,\bR} &= -S + \hat{b}^{\dag}_{\bR}\hat{b}_{\bR},
  & \hat{S}^{+}_{B,\bR} &\simeq \sqrt{2S}\,\hat{b}^{\dag}_{\bR},
  & \hat{S}^{-}_{B,\bR} &\simeq \sqrt{2S}\,\hat{b}_{\bR}.
\end{align}
Fourier transformation, $\hat{a}_{\bR}=N^{-1/2}\sum_{\bk}\hat{a}_{\bk}e^{i\bk\cdot\bR}$ and similarly for $\hat{b}$, with $N$ the number of unit cells, casts the bilinear part of $\hat{H}_{\rm spin}$ into the standard antiferromagnetic spin-wave form, diagonalized by a Bogoliubov rotation with magnon branches $\mu$ and dispersions $\epsilon_{\bk,\mu}$.

The linearized transformation~\eqref{eqn:HP_SI} is controlled by the magnon dilution $\langle\hat{a}^{\dag}\hat{a}\rangle/2S\ll 1$. The vertices derived below are therefore zero-temperature objects. They enter the analysis only through the kinematic densities of Sec.~1.7; the strongly heated regime, where the dilution parameter is not small, is treated nonperturbatively by the coupled spin-lattice simulations of the main text.

\subsection{Phonons}

The displacement of atom $n$ (mass $m_{n}$) in the unit cell at $\bR$ is expanded over phonon modes $(\bq,\lambda)$ with frequency $\Omega_{\bq,\lambda}$, polarization $\bepsilon^{n}_{\bq,\lambda}$, and ladder operators $\hat{c}_{\bq,\lambda}^{(\dag)}$:
\begin{equation}\label{eqn:phonon_expansion_SI}
  \bu_{n,\bR}
  = \sum_{\bq,\lambda}
    \sqrt{\frac{\hbar}{2 N m_{n} \Omega_{\bq,\lambda}}}\;
    \bepsilon^{n}_{\bq,\lambda}
    \bigl(\hat{c}_{\bq,\lambda}\,e^{i\bq\cdot\bR}
        + \hat{c}_{\bq,\lambda}^{\dag}\,e^{-i\bq\cdot\bR}\bigr),
\end{equation}
with the polarization vectors normalized as $\sum_{n}|\bepsilon^{n}_{\bq,\lambda}|^{2}=1$.

\subsection{Exchange striction and the one-phonon-two-magnon vertex}

The exchange constant of a bond depends on the positions of all atoms that contribute to its orbital overlaps, including non-magnetic bridging atoms. To leading order in the displacements,
\begin{equation}\label{eqn:J_linear_SI}
  J_{ij}\bigl(\{\bu\}\bigr)
  = J^{(0)}_{ij}
  + \sum_{n,\bR'} \frac{\partial J_{ij}}{\partial \bu_{n,\bR'}}\cdot\bu_{n,\bR'}
  + O(u^{2}).
\end{equation}
The derivatives fall off rapidly with the distance between the displaced atom and the bond, and only a few atoms contribute (Fig.~2c of the main text). By translational invariance the derivative depends on relative coordinates only. We group the bonds by the vector $\bdelta$ that connects an $A$ site at $\bR$ to its $B$ partner at $\bR+\bdelta$ and define the striction tensor $\bK^{n}_{\bdelta}(\bR') \equiv \partial J_{AB,\bdelta}/\partial\bu_{n,\bR'}$ for the reference bond at $\bR=0$. Grouping the double sum of Eq.~\eqref{eqn:Hspin_SI} into $A$-to-$B$ bonds absorbs the factor $1/2$, and each bond appears once:
\begin{equation}\label{eqn:Hint_ex_real_SI}
  \hat{H}_{\rm int}^{\rm ex}
  = \sum_{\bR}\sum_{\bdelta}\sum_{n,\bR'}
    \bigl[\bK^{n}_{\bdelta}(\bR')\cdot\bu_{n,\bR+\bR'}\bigr]\,
    \hat{\bS}_{A,\bR}\cdot\hat{\bS}_{B,\bR+\bdelta}.
\end{equation}

The HP expansion of the spin product reads
\begin{equation}\label{eqn:spin_product_SI}
  \hat{\bS}_{A,\bR}\cdot\hat{\bS}_{B,\bR+\bdelta}
  = -S^{2}
  + S\bigl(\hat{a}^{\dag}_{\bR}\hat{a}_{\bR}
         + \hat{b}^{\dag}_{\bR+\bdelta}\hat{b}_{\bR+\bdelta}
         + \hat{a}_{\bR}\hat{b}_{\bR+\bdelta}
         + \hat{a}^{\dag}_{\bR}\hat{b}^{\dag}_{\bR+\bdelta}\bigr)
  + O(S^{0}).
\end{equation}
The three groups of terms play different roles. The constant $-S^{2}$ multiplies $\bK\cdot\bu$ and is a static force; it renormalizes the reference structure and produces no magnon dynamics. The diagonal terms $\hat{a}^{\dag}\hat{a}$ and $\hat{b}^{\dag}\hat{b}$ contribute to magnon self-energies but conserve the magnon number. The pair term $\hat{a}\hat{b}+\hat{a}^{\dag}\hat{b}^{\dag}$ creates or annihilates a magnon pair on the bond; combined with the displacement, which is linear in the phonon operators, it drives a one-phonon-two-magnon (1p2m) process. Substituting Eqs.~\eqref{eqn:HP_SI} and~\eqref{eqn:phonon_expansion_SI} into Eq.~\eqref{eqn:Hint_ex_real_SI} and Fourier transforming gives
\begin{equation}\label{eqn:Hint_ex_Fourier_SI}
  \hat{H}_{\rm int}^{\rm ex}
  = \frac{1}{\sqrt{N}}\sum_{\bq,\lambda}\sum_{\bk}
    V^{\lambda}_{\bk,\bq}\,
    \bigl(
      \hat{c}_{\bq,\lambda}\,\hat{a}_{\bk}^{\dag}\hat{b}_{\bq-\bk}^{\dag}
      + \hat{c}_{\bq,\lambda}^{\dag}\,\hat{a}_{\bk}\hat{b}_{\bq-\bk}
    \bigr)
    \;+\;\hat{H}_{\rm diag},
\end{equation}
with the vertex
\begin{equation}\label{eqn:V_ex_SI}
  V^{\lambda}_{\bk,\bq}
  = S\sum_{\bdelta}\sum_{n,\bR'}
    \sqrt{\frac{\hbar}{2 m_{n} \Omega_{\bq,\lambda}}}\;
    \bigl(\bK^{n}_{\bdelta}(\bR')\cdot\bepsilon^{n}_{\bq,\lambda}\bigr)\,
    e^{i\bq\cdot\bR'}\,e^{-i(\bq-\bk)\cdot\bdelta},
\end{equation}
where $\hat{H}_{\rm diag}$ collects the diagonal contributions. The two phase factors have distinct origins: $e^{i\bq\cdot\bR'}$ tracks the position of the displaced atom relative to the bond, and $e^{-i(\bq-\bk)\cdot\bdelta}$ tracks the position of the $B$ partner, $\bq-\bk$ being the momentum of the magnon created on sublattice $B$. The Bogoliubov rotation dresses the vertex with coherence factors of order unity; they drop out of the kinematic densities used below and are comparable in the two compounds, so we do not carry them explicitly.

\subsection{Fermi's golden rule rate}

The matrix element of Eq.~\eqref{eqn:Hint_ex_Fourier_SI} between a one-phonon state and a two-magnon state is $V^{\lambda}_{\bk,\bq}/\sqrt{N}$, and the final states are labelled by the internal momentum $\bk$. Fermi's golden rule then gives the phonon decay rate
\begin{equation}\label{eqn:golden_rule_SI}
  \tau_{\bq,\lambda}^{-1}
  = \frac{2\pi}{\hbar N} \sum_{\bk,\mu,\nu}
    |V^{\lambda}_{\bk,\bq}|^{2}\;
    \delta\bigl(\hbar\Omega_{\bq,\lambda}
          - \epsilon_{\bk,\mu} - \epsilon_{\bq-\bk,\nu}\bigr).
\end{equation}
The factor $1/N$ combines with the sum over $\bk$ into a Brillouin-zone average, $N^{-1}\sum_{\bk} \to v_{c}\int d^{3}k/(2\pi)^{3}$ with $v_{c}$ the unit-cell volume, so the rate is intensive. Equation~\eqref{eqn:golden_rule_SI} is the rate used in Eq.~(1) of the main text.

\subsection{Dzyaloshinskii-Moriya channels}

The DM term of Eq.~\eqref{eqn:Hspin_SI} has a different operator structure. Its transverse components pair one transverse spin operator, of order $\sqrt{S}$, with one longitudinal one, of order $S$, and therefore contain terms linear in the magnon operators. Combined with the displacement expansion of $\bD_{ij}$, these produce a one-phonon-one-magnon (1p1m) hybridization,
\begin{equation}\label{eqn:Hint_DM_SI}
  \hat{H}_{\rm int}^{\rm DM}
  = \sum_{\bq,\lambda}
    W^{\lambda}_{\bq}\,
    \bigl(\hat{c}_{\bq,\lambda}\,\hat{a}_{\bq}^{\dag}
        + \hat{c}_{\bq,\lambda}^{\dag}\,\hat{a}_{\bq}\bigr)
    \;+\;\text{(1p2m terms)},
\end{equation}
with a vertex of magnitude
\begin{equation}\label{eqn:W_DM_SI}
  |W^{\lambda}_{\bq}|
  \sim S^{3/2}
    \sqrt{\frac{\hbar}{2 \bar{m}\, \Omega_{\bq,\lambda}}}\;
    \Bigl|\frac{\partial\bD}{\partial\bu}\Bigr|,
\end{equation}
where $\bar{m}$ is a representative atomic mass. The hybridization term carries no factor $1/\sqrt{N}$, because the lattice sum enforces equal phonon and magnon momenta, and its golden-rule rate requires a phonon-magnon band crossing at the same $\bq$:
\begin{equation}\label{eqn:golden_rule_DM_SI}
  \bigl(\tau^{\rm DM}_{\bq,\lambda}\bigr)^{-1}
  = \frac{2\pi}{\hbar} \sum_{\mu}
    |W^{\lambda}_{\bq}|^{2}\,
    \delta\bigl(\hbar\Omega_{\bq,\lambda} - \epsilon_{\bq,\mu}\bigr).
\end{equation}
The longitudinal component of the cross product pairs two transverse operators and adds a 1p2m term of order $S$ with the same kinematic structure as Eq.~\eqref{eqn:V_ex_SI}, with $\bK \to \partial D^{z}/\partial\bu$.

We do not need the explicit DM vertices. Assuming that the strictive coupling of each interaction scales with its equilibrium magnitude, $|\partial\bD/\partial\bu| \sim (D/J)\,|\partial J/\partial\bu|$, both DM channels are suppressed by $(D/J)^{2}$ at the matrix-element level relative to the exchange-driven 1p2m channel. In FeBO$_{3}$ the $1^{\circ}$ canting~\cite{eibschutz_1970,dmitrienko_2014} gives $D/J\approx 0.04$ and $(D/J)^{2}\approx 1.6\times 10^{-3}$; in Cr$_{2}$O$_{3}$ the ratio is smaller still. Displacements can also induce DM components on bonds where the equilibrium $\bD$ is suppressed by symmetry rather than by weak spin-orbit coupling; such contributions are bounded by the spin-orbit ratio $\lambda/\Delta \approx 0.005$--$0.01$ for the orbital-singlet ground state of Cr$^{3+}$, while for FeBO$_{3}$ the ratio $D/J$ is measured directly through the canting, so the estimate below is unchanged. The exchange channel itself generates no 1p1m term, because $\hat{\bS}_{i}\cdot\hat{\bS}_{j}$ contains no contribution linear in the magnon operators.

\subsection{Kinematic phase-space densities}

The kinematic content of Eqs.~\eqref{eqn:golden_rule_SI} and~\eqref{eqn:golden_rule_DM_SI} is evaluated from the \emph{ab initio} dispersions on a uniform $N_{g}\times N_{g}\times N_{g}$ grid spanning the rhombohedral Brillouin zone ($N_{g}=50$, reduced coordinates in $[0,1)^{3}$). The energy $\delta$-functions are regularized by a normalized Gaussian $G_{\sigma}$ of width $\sigma=0.25$~THz, small compared with the phonon and magnon bandwidths but large enough to bridge the energy spacing between adjacent grid points. Varying $\sigma$ between 0.1 and 1~THz changes the absolute densities but leaves the cross-material and cross-channel ratios unchanged. The two densities are
\begin{equation}\label{eqn:D_1p1m_SI}
  D^{(1p1m)}(\bq)
  = \sum_{\lambda,\mu}
    G_{\sigma}\!\bigl(\hbar\Omega_{{\rm ph},\lambda}(\bq)
                    - \epsilon_{{\rm mag},\mu}(\bq)\bigr),
\end{equation}
\begin{equation}\label{eqn:D_1p2m_SI}
  D^{(1p2m)}(\bq)
  = \sum_{\bk}\sum_{\lambda,\mu,\nu}
    G_{\sigma}\!\bigl(\hbar\Omega_{{\rm ph},\lambda}(\bq)
                    - \epsilon_{{\rm mag},\mu}(\bk)
                    - \epsilon_{{\rm mag},\nu}(\bq-\bk)\bigr),
\end{equation}
shown, as raw sums, across the full Brillouin zone in Figures~S3 and~S4. Momentum conservation in Eq.~\eqref{eqn:D_1p2m_SI} is enforced by index arithmetic on the grid, with wraparound across the zone boundaries automatically absorbing Umklapp contributions. $D^{(1p1m)}$ is intensive, while $D^{(1p2m)}$ contains the unrestricted internal sum over $\bk$ and scales with the number of grid points; the corresponding intensive quantity, entering the rate~\eqref{eqn:golden_rule_SI}, is $N_{g}^{-3} D^{(1p2m)}$. All comparisons between the two compounds are made at equal $N_{g}$.

\subsection{Channel hierarchy and cross-material comparison}

Three channels compete: exchange-driven 1p2m, DM-driven 1p2m, and DM-driven 1p1m. The exchange-driven 1p2m process sets the reference rate. The DM-driven 1p2m process shares its kinematics but carries the matrix-element suppression $(D/J)^{2}\lesssim 10^{-3}$. The DM-driven 1p1m process carries the same suppression and is in addition kinematically confined to phonon-magnon band crossings, which form a measure-zero set on the dispersions, whereas the 1p2m resonance condition leaves the internal momentum free. With equal-magnitude striction assumed for both interactions, the rate ratio is
\begin{equation}\label{eqn:hierarchy_SI}
  \frac{(\tau^{\rm DM,1p1m})^{-1}}{(\tau^{\rm ex,1p2m})^{-1}}
  \sim \Bigl(\frac{D}{J}\Bigr)^{2}\,
       \frac{D^{(1p1m)}}{N_{g}^{-3}\,D^{(1p2m)}}.
\end{equation}
The Brillouin-zone averages of the computed densities are $\langle D^{(1p1m)}\rangle = 6.8$ and $2.6~\mathrm{THz}^{-1}$, and $N_{g}^{-3}\langle D^{(1p2m)}\rangle = 27.5$ and $8.6~\mathrm{THz}^{-1}$, for Cr$_{2}$O$_{3}$ and FeBO$_{3}$, respectively. The intensive 1p2m density therefore exceeds the 1p1m density by a factor of 3--4 in both compounds, and the ratio~\eqref{eqn:hierarchy_SI} stays below $5\times 10^{-4}$. All DM contributions are negligible, and only the exchange-driven 1p2m channel is retained in the analysis of the main text.

The cross-material asymmetry in the relaxation rate traces back to two factors. The striction tensor $\bK^{n}_{\bdelta}(\bR')$ is roughly an order of magnitude larger in Cr$_{2}$O$_{3}$ than in FeBO$_{3}$, because the direct $t_{2g}$-$t_{2g}$ overlap on the short Cr-Cr bonds depends exponentially on the cation-cation distance. The Brillouin-zone-integrated 1p2m density is about three times larger in Cr$_{2}$O$_{3}$, 27.5 against 8.6~THz$^{-1}$ in the intensive normalization, because its wide magnon band produces a two-magnon continuum that overlaps nearly the entire phonon spectrum. Both factors, and the resulting rate estimate, are discussed in the main text.

\clearpage

\section*{Supporting Figures}

\begin{figure}[htbp]
\centering
\includegraphics[width=\linewidth]{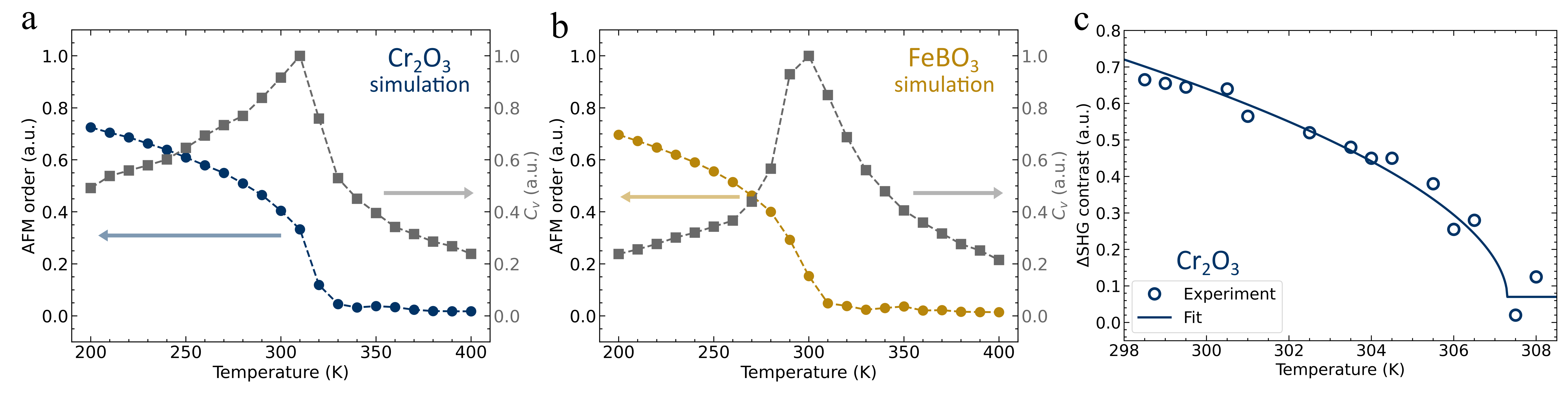}
\caption{\textbf{Temperature dependence of the antiferromagnetic order parameter.}
(a,b) AFM order parameter (circles) and specific heat (squares) versus temperature from UppASD simulations for (a) Cr$_2$O$_3$ and (b) FeBO$_3$.
(c) Experimental temperature dependence of the SHG contrast $\Delta \text{SHG}$ extracted from SHG snapshots (symbols), fitted with $A(T_N-T)^\beta + y_0$ (line), yielding $T_N = 307.3\text{~K}$ and $\beta = 0.54 \pm 0.08$.}
\label{ed:AFM_vs_T}
\end{figure}

\begin{figure}[htbp]
\centering
\includegraphics[width=0.75\linewidth]{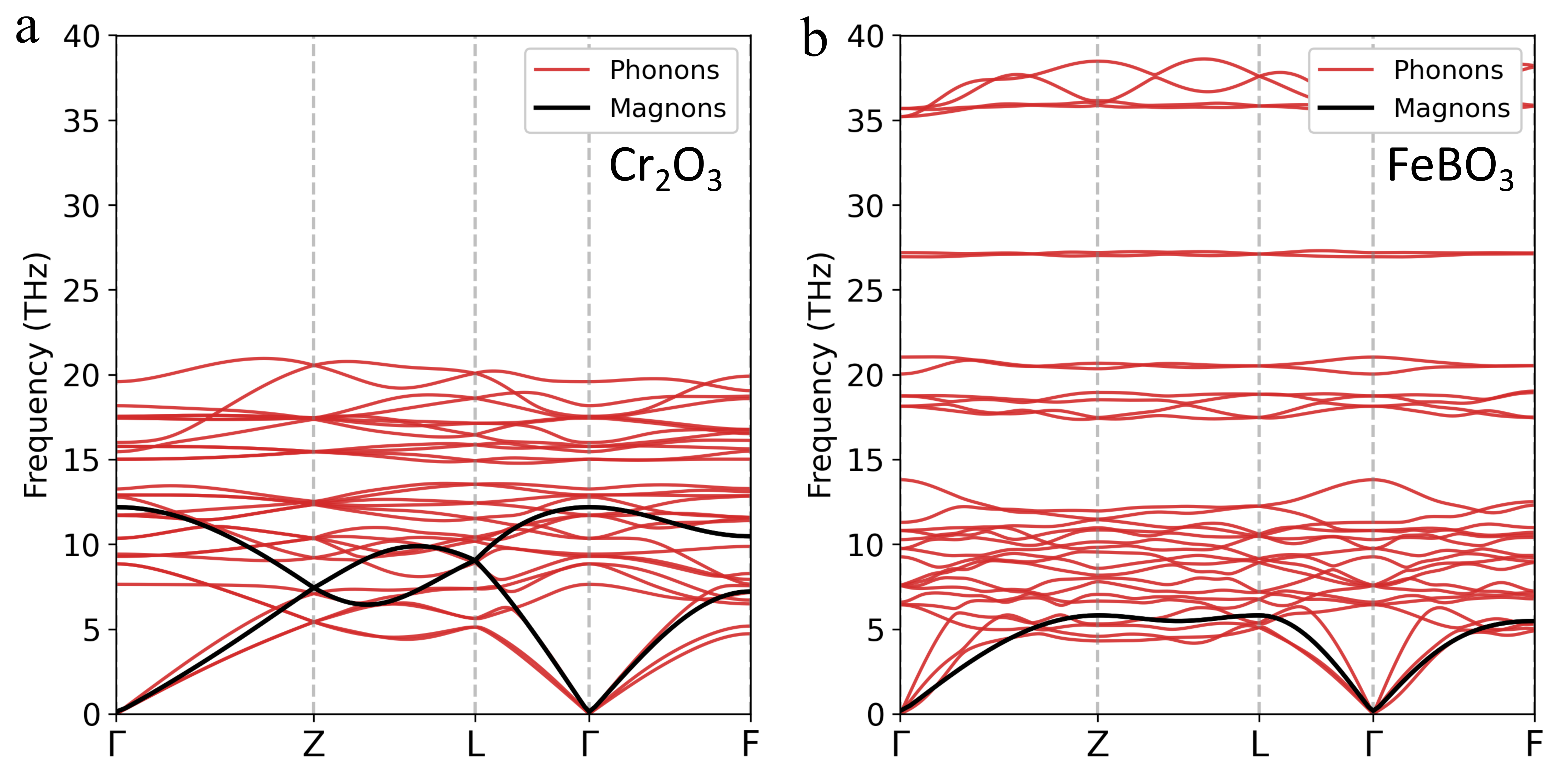}
\caption{\textbf{Phonon and magnon dispersions in Cr$_2$O$_3$ and FeBO$_3$.}
Phonon (red) and magnon (black) dispersions along the high-symmetry path in the Brillouin zone for (a) Cr$_2$O$_3$ and (b) FeBO$_3$.}
\label{ed:magnons_phonons_dispersions}
\end{figure}

\begin{figure}[htbp]
\centering
\includegraphics[width=\linewidth]{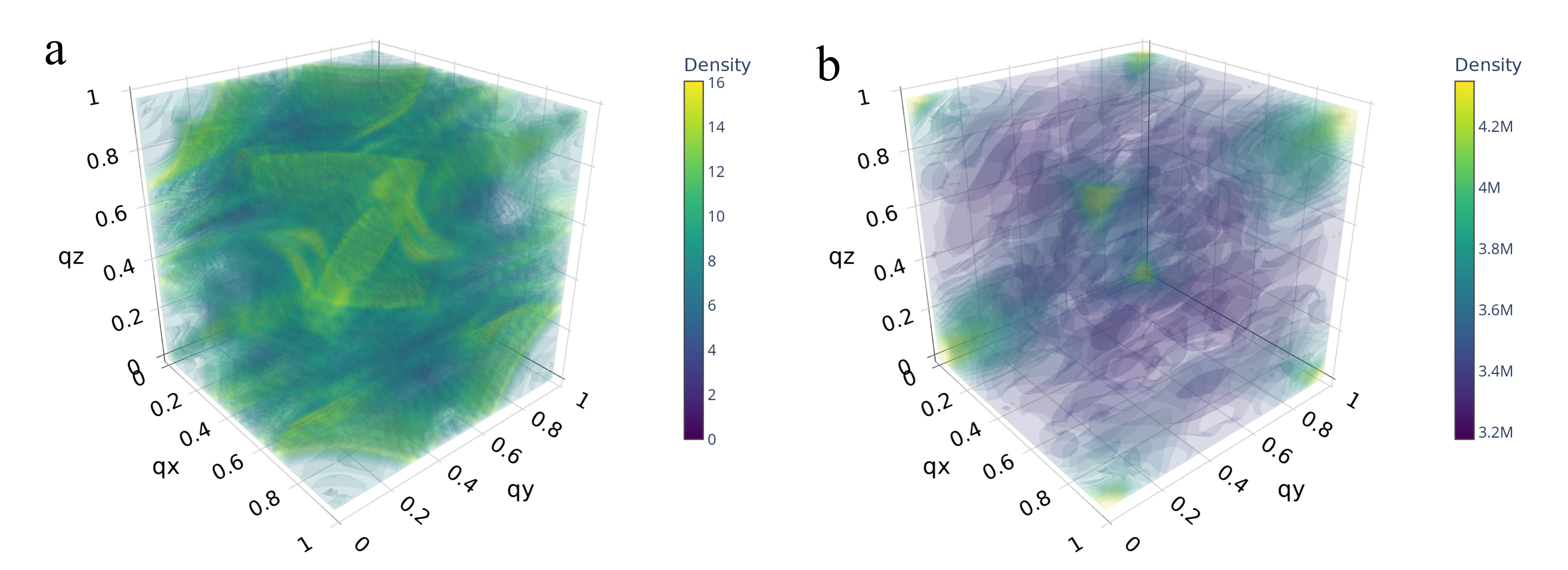}
\caption{\textbf{Kinematic phase spaces for Cr$_2$O$_3$.}
$D^{(1p1m)}(\bq)$ (a) and $D^{(1p2m)}(\bq)$ (b) across the full Brillouin zone, obtained from the \emph{ab initio} phonon and magnon dispersions with unit matrix elements as described in Methods.}
\label{ed:phase_space_Cr2O3}
\end{figure}

\begin{figure}[htbp]
\centering
\includegraphics[width=\linewidth]{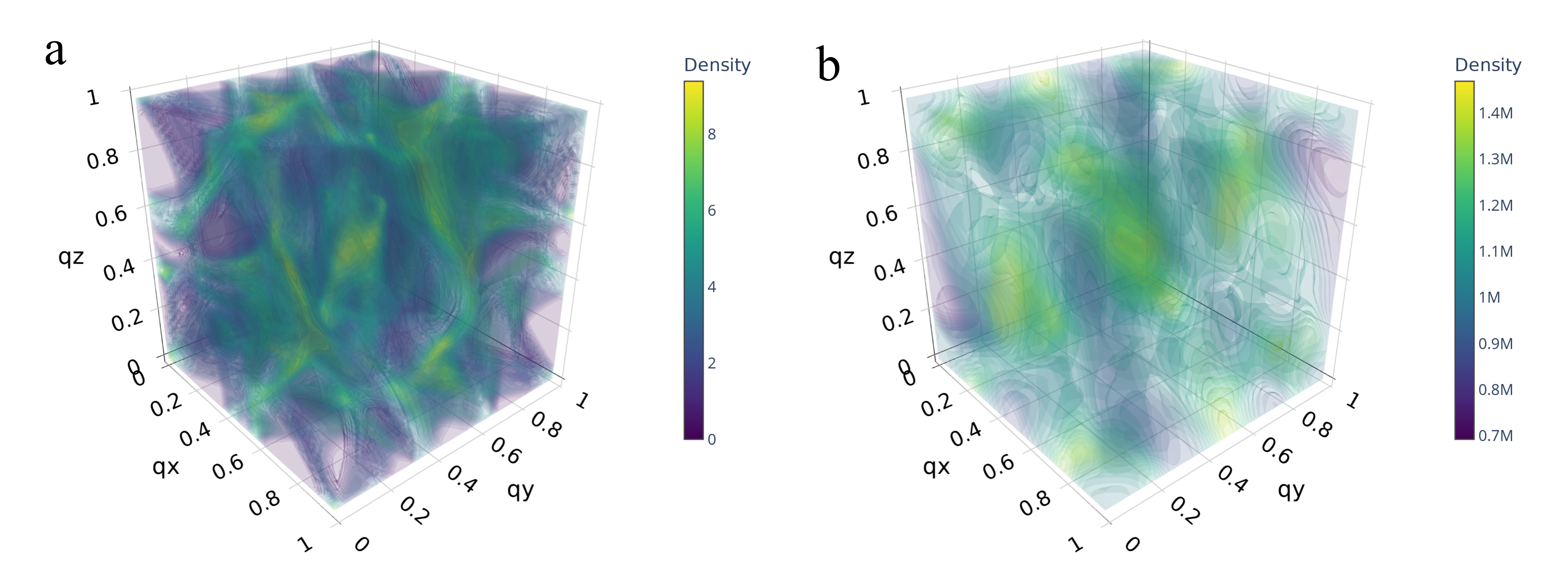}
\caption{\textbf{Kinematic phase spaces for FeBO$_3$.}
Same as Fig.~\ref{ed:phase_space_Cr2O3}, computed for FeBO$_3$.}
\label{ed:phase_space_FeBO3}
\end{figure}

\begin{figure}[htbp]
\centering
\includegraphics[width=1.0\linewidth]{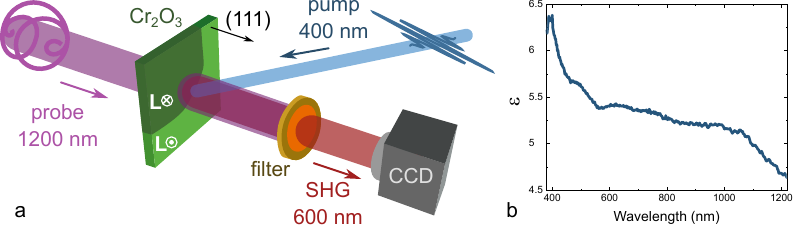}
\caption{\textbf{Experimental setup and dielectric properties of Cr$_2$O$_3$.}
(a) Schematic of the experimental setup for time-resolved second-harmonic generation (SHG) imaging. The Cr$_2$O$_3$ sample is excited by 400~nm pump pulses and probed by circularly polarized 1200~nm pulses. The resulting 600~nm SHG signal from the probe pulse is filtered and imaged onto a CCD camera to visualize the AFM domains with opposite orientations of the N\'eel vector $\mathbf{L}$.
(b) Measured spectrum of dielectric permittivity of the crystal.}
\label{ed:expr_setup}
\end{figure}

\end{document}